# Negative Friction Coefficient in Superlubric Graphite/Hexagonal Boron Nitride Heterojunctions

Davide Mandelli, Wengen Ouyang, Oded Hod,* Michael Urbakh

Department of Physical Chemistry, School of Chemistry, The Raymond and Beverly Sackler Faculty of Exact Sciences and The Sackler Center for Computational Molecular and Materials Science, Tel Aviv University, Tel Aviv 6997801, Israel.

## Abstract

Negative friction coefficient, where friction is reduced upon increasing normal load, is predicted for superlubric graphite/hexagonal boron nitride heterojunctions. The origin of this counterintuitive behavior lies in the load-induced suppression of the moiré superstructure out-of-plane distortions leading to a less dissipative interfacial dynamics. Thermal induced enhancement of the out-of-plane fluctuations leads to unusual increase of friction with temperature. The highlighted frictional mechanism is of general nature and is expected to appear in many layered materials heterojunctions.

Energy dissipation, wear, and the ensuing failure of moving components are problems encountered in many human activities. Reducing friction is of particular importance in microscopic and nanoscopic mechanical devices [1], where the damaging consequences of local heating are amplified by the large surface-to-volume ratio. Standard lubrication schemes may fail in these extremely confined conditions, which calls for novel alternative solutions [2], such as dry solid coatings. In this approach, frictional forces are reduced thanks to the effective cancellation of lateral interactions occurring between incommensurate rigid crystalline surfaces. This phenomenon, often termed structural superlubricity, was first proposed theoretically a few decades ago [3] as a way to achieve extremely low friction coefficients [4–7]. Despite the fact that structural superlubricity has been observed in different material contacts [8–12], its implementation in practical solid/solid lubrication schemes remains a challenging task. Nevertheless, recent breakthrough demonstrations [13–17], based on van-der-Waals heterojunctions, suggest that this goal may be within our reach.

Among the family of layered compounds, junctions formed between graphite and hexagonal boron nitride (h-BN) have been predicted as promising candidates to achieve robust superlubricity [18,19].

This was recently verified experimentally for micron-sized monocrystalline interfaces [16], where superlubricity was found to persist even under ambient conditions and over a broad range of sliding velocities. However, due to experimental restrictions, measurements were performed only at relatively small applied normal pressures of $\lesssim 10$ MPa.

In this letter, we use molecular dynamics simulations to investigate the frictional response of extended graphite/$h$-BN contacts under uniform normal loads up to $\sim 10$ GPa. Using state-of-the-art interlayer force-fields we show that superlubricity persists across the entire range of normal loads investigated. Surprisingly, we find that kinetic friction displays a non-monotonic behavior with external load, where an initial regime of friction reduction (by up to $\sim 30$ % relative to the zero-load value) is followed by friction increase for higher normal loads. These negative friction coefficients (NFCs) originate from the load-induced suppression of energy dissipation via out-of-plane atomic motion. This behavior is expected to appear in many heterogeneous two-dimensional material interfaces that possess a corrugated moiré superstructure and exhibit superlubric motion.

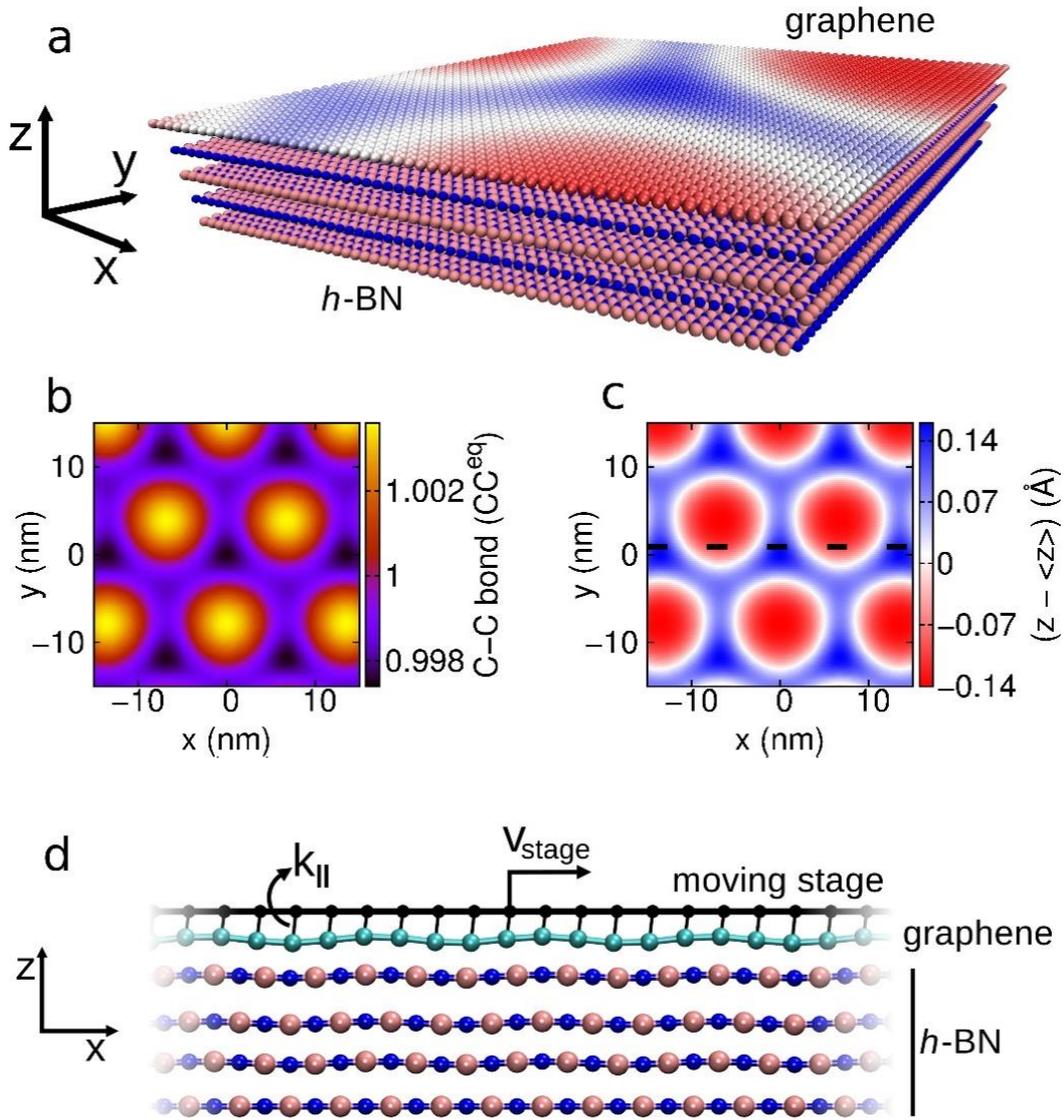

**Figure 1**: (a) The model heterojunction: a graphene monolayer aligned over a four-layers thick *h*-BN substrate with rigid bottom layer. Carbon atoms are colored according to their vertical position with respect to the average basal plane of the graphene layer, where blue and red correspond to the maximum and minimum values attained, respectively. The scale bar is the same as in panel c. (b) Colored map of the average carbon-carbon distance (in units of the equilibrium bond-length) after relaxation of the graphene layer over *h*-BN at zero normal load. (c) The corresponding colored map of the out-of-plane distortions of graphene. (d) Schematics of the simulations setup: the graphene layer is attached to a rigid stage moving at fixed velocity of $v_{stage}$ along the *x* direction. Driving forces are exerted via identical springs of elastic constant $k_\parallel$ acting only parallel to the substrate. No forces are exerted by the springs in the normal direction. The perpendicular position of the stage is shifted vertically for clarity of the presentation. Carbon, boron, and nitrogen atoms are represented by cyan, pink, and blue spheres, respectively.

Our main goal is to investigate the frictional response of layered graphene/*h*-BN heterojunctions over a broad range of normal loads and identify conditions where NFCs can be obtained. To this end, we consider the model system presented in Fig. 1a. Here, a laterally periodic graphene monolayer is aligned over a substrate made of four *h*-BN layers, with parallel crystalline axes. The carbon-carbon and boron-nitrogen intralayer interactions are modeled via the REBO [20,21] and the Tersoff [22] potentials, respectively. The interlayer interactions are described via the graphene/*h*-BN interlayer potential [23,24]. This model describes well the frictional behavior of realistic three dimensional graphite/*h*-BN contacts [16]. Due to the inherent 1.8% lattice constant mismatch between graphene and *h*-BN, a moiré superstructure of period $\lambda_\mathrm{m} \approx 14$ nm appears in the fully-relaxed interface. In agreement with experiments [25], we observe a periodic sequence of extended domains, where graphene remains flat and stretches to reach local lattice commensurability with the underlying substrate, separated by narrow domain-walls, where in-plane compressive strain is accumulated (see Fig. 1b). The latter is partially relieved via the formation of elevated ridges (see Fig. 1c), which minimize the total energy by converting energetically costly in-plane compressions into cheaper local out-of-plane distortions (see section 1 of Supplemental Material (SM) [26] for details). This structure may present friction coefficients well within the superlubric regime [15,16,18] due to the soliton-like motion [31] of the elevated moiré ridges [19].

To model the effects of external load on friction, we perform fully atomistic dynamic simulations, where the graphene layer is attached to a stage moving at constant velocity (see Fig. 1d), mimicking recent sliding friction experiments of microscale graphite/*h*-BN contacts [16]. The stage is modeled as a rigid flat duplicate of the graphene slider and the normal load is simulated by applying a uniform force to all slider atoms, vertical to the graphene basal plane and pressing towards the *h*-BN substrate. In realistic scenarios this can be achieved by a sufficiently thick slider, which helps spreading the load evenly across the interface [16]. The equations of motion for the carbon atoms are given by:

$$m_\mathrm{C}\ddot{\boldsymbol{r}}_i = -\boldsymbol{\nabla}_{\boldsymbol{r}_i}(V_\mathrm{inter} + V_\mathrm{intra}) + k_\parallel \left(\boldsymbol{r}_{\parallel,i}^\mathrm{stage} - \boldsymbol{r}_{\parallel,i}\right) - F_\mathrm{N}\hat{z} - m_C \sum_{\alpha=x,y,z} \eta_\alpha v_{\alpha,i}\,\hat{\boldsymbol{\alpha}}, \qquad (1)$$

where the first two terms on the right-hand-side are the forces due to the inter- and intralayer interactions, the third term is the lateral elastic driving force due to the moving stage, the fourth term is the applied normal load, and the last term is a viscous force that accounts for energy dissipation, needed to reach steady-state [32]. Similar equations are solved for the motion of the boron and nitrogen atoms within the mobile $h$-BN substrate layers, with the absence of the driving term and the normal load. All atoms are free to move in any directions, apart from those of the bottommost $h$-BN layer, which are held fixed at their equilibrium positions, and those of the moving stage that are rigidly shifted in the $x$ direction at a constant velocity. The anisotropic nature of the system is expected to affect the rate of kinetic energy dissipation in different directions. Previous studies predicted that the vertical damping coefficient can be much larger than the lateral ones [33,34], hence we fixed $\eta_x = \eta_y = 0.1 \text{ ps}^{-1}$ and $\eta_z = 4.5 \text{ ps}^{-1}$. These parameters, which fall within the typical range used in molecular dynamics simulations of nanoscale friction [32], provide good agreement with experimental data on the anisotropy of superlubric graphite/$h$-BN heterojunctions [16] and are close to theoretical estimations for atomic adsorbates [35]. The kinetic frictional stress, $\sigma_k$, is evaluated as the time-average of the total shear-force acting on the moving stage in the sliding direction, normalized to contact area $A$,

$$\sigma_k = \frac{\langle \sum_{i=1}^{N_C} k_\parallel (x_i^{\text{stage}} - x_i) \rangle}{A}. \tag{2}$$

Here, $N_C$ is the number of slider carbon atoms, $x_i$ and $x_i^{\text{stage}}$ are the positions of the $i$-th carbon atom and of its counterpart in the moving stage, respectively, and $k_\parallel = 11 \text{ meV/Å}^2$ is the spring constant (see further simulation details in SM [26] sections 2 and 3).

The main results of our study are presented in Fig. 2, showing non-monotonic variation of the frictional stress with respect to the applied normal load at zero and room temperature, resulting in NFCs. Considering first the zero temperature results (red curve) we find that up to an external pressure of ~5 GPa the frictional stress steadily reduces by ~20 % from ~2.4 MPa down to ~1.9 MPa. Further increase of the external load results in an increase in the frictional stress up to ~2.1 MPa obtained at the highest load investigated of ~12 GPa. This latter observation is in qualitative agreement with recent experimental results showing linear increase of friction of graphene coated microspheres [15] or graphite flake wrapped AFM tips [17] sliding atop $h$-BN.

The corresponding differential friction coefficient are obtained by performing a numerical two-point derivative of the data yielding kinetic friction coefficients in the range of $-2.5 \times 10^{-4} < \mu_k < 5 \times 10^{-5}$, well within the superlubric regime, defined by $\mu_k < 10^{-3}$ [2]. Simulations performed at room temperature (black curves) yielded a similar frictional stress reduction of $27 \pm 5$ % relative to the zero-load value (see section 4 of SM [26] for further details). In striking difference from standard frictional scenarios, where friction reduces with increasing temperature [32], in the present case friction is found to enhance with temperature. We note that while our simulations are performed under uniform normal load, similar effects are expected to appear also for non-uniform load distributions as long as the local load does not exceed the turnover value, above which kinetic friction starts to increase.

To evaluate the effect of slider thickness, we have performed geometry optimizations of a stack of six graphene layers atop six *h*-BN layers under external load and found a similar load-dependence of the moiré vertical distortions (see SM [26] section 5), thus indicating the validity of the graphene monolayer model adopted herein. Furthermore, we expect that relaxing the periodic boundary conditions, valid for large-scale monocrystalline interfaces [16], will allow for more pronounced load induced flattening of the moiré ridges resulting in larger absolute NFCs.

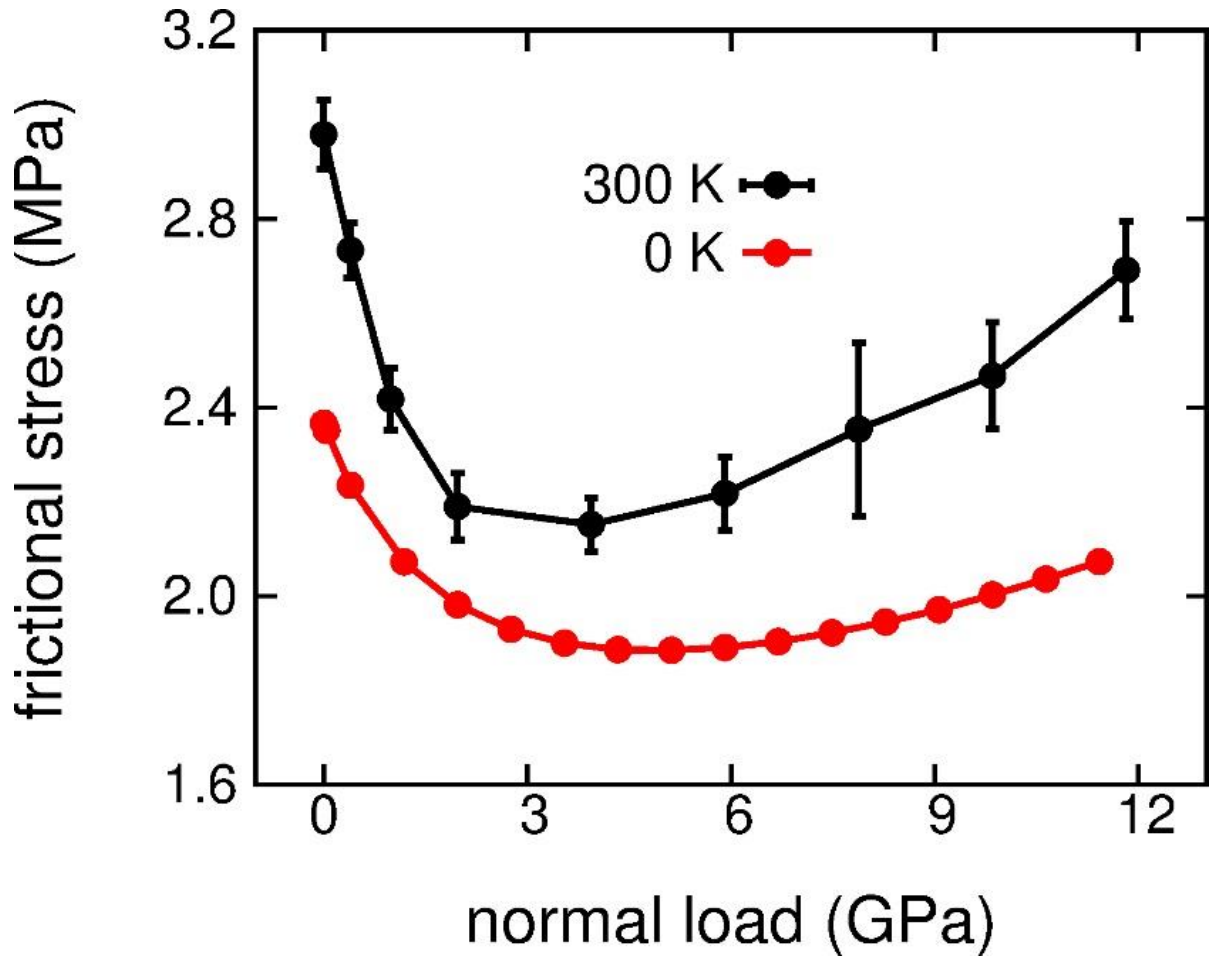

**Figure 2**: Load dependence of the frictional stress obtained at zero (red) and room (black) temperature. Standard error bars are computed by block averaging.

The mechanism underlying this counterintuitive behavior can be unveiled by investigating the dependence of the different energy dissipation routes on the applied normal load. Hence, we calculate the steady-state frictional power, $p_{\text{fric}}$, dissipated at zero temperature, from which the kinetic frictional stress can be evaluated as follows [36]:

$$\sigma_k = p_{\text{fric}}/Av_{\text{stage}} = (p_{\text{com}} + p_{\text{internal}})/Av_{\text{stage}}. \tag{3}$$

Here,

$$p_{\text{com}} = \sum_{i=1}^{N_{\text{layers}}} M_i \sum_{\alpha=x,y,z} \eta_\alpha \langle \left(v^i_{\alpha,\text{com}}(t)\right)^2 \rangle \tag{4}$$

is the time-averaged power dissipated by the center-of-mass degree-of-freedom of each layer, where $N_{\text{layers}}$ is the total number of layers, $M_i$ is the total mass of the $i$-th layer, and $v^i_{\alpha,\text{com}}$ is its center-of-mass velocity along Cartesian direction $\alpha = x, y, z$. $p_{\text{internal}}$ accounts for the power dissipated through the motion of the internal degrees-of-freedom in the center-of-mass reference frame, and is given by:

$$p_{\text{internal}} = \sum_{i=1}^{N_{\text{layers}}} \sum_{k=1}^{N_i} m^i_k \sum_{\alpha=x,y,z} \left[\eta_\alpha \langle \left(v^i_{\alpha,k}(t) - v^i_{\alpha,\text{com}}(t)\right)^2 \rangle\right], \tag{5}$$

where $N_i$ is the number of atoms in the $i$-th layer and $m^i_k$ and $v^i_{\alpha,k}$ are the mass and $\alpha$ Cartesian velocity component of the $k$-th atom in that layer, respectively.

Figure 3a reports the center-of-mass contribution to dissipation obtained at 0 K, which is dominated by its component along the sliding direction (black curve in Fig. 3a). Notably, the center-of-mass dissipation shows no significant load dependence suggesting that the origin of the NFCs reported above lies within the dynamics of the internal degrees-of-freedom of the slider. To verify this, we show in Fig. 3b the overall contributions from the internal degrees-of-freedom of all layers at zero temperature. We observe that this is dominated by the vertical component (red curve in Figs. 3b), which displays the reported non-monotonic trend. We further note that over 90% of the energy dissipated is associated with the motion of the upper graphene layer (see SM [26] section 6). Therefore, the non-monotonic frictional behavior stems from the effect of normal load on the out-of-plane motion of the graphene atoms.

Performing similar analysis at room temperature is more delicate, since the statistical errors in the frictional power are much larger than the overall average frictional stress. Nevertheless, the similar trends obtained for the frictional stress at zero and room temperatures (see Fig. 2) suggest that the same physical mechanism underlies the observed non-monotonic behavior in both cases. Furthermore, the thermally induced enhancement of the out-of-plane fluctuations explains the unusual increase of friction with temperature.

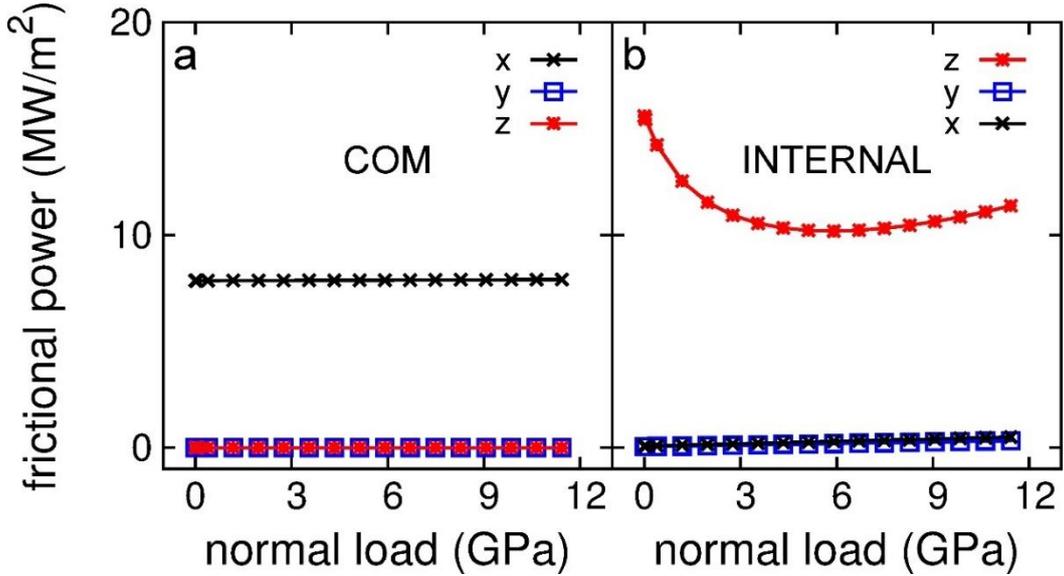

**Figure 3**: Load dependence of the frictional power dissipated at zero temperature by (a) the center-of-mass and (b) the internal degrees-of-freedom of all model heterojunction mobile layers.

The key to understand the obtained NFCs is to study the dependence of the moiré ridge structure and its soliton-like dynamics on the applied external load [16,19]. In Fig. 4a we present the ridge profile as obtained at three normal load values: zero load, ~ 4 GPa (where the friction stress obtains a minimum), and ~11 GPa. The profile is taken along a scanline in the sliding direction that corresponds to the maximum peak-dip value of the vertical deformations of graphene (see dashed black line in Fig. 1c). We observe a strong initial decrease of the ridge peak height accompanied by its narrowing when the load increases up to 4 GPa (see also Fig. 4b). Further increasing the pressure to ~11 GPa results in merely mild reduction of the ridge peak height with additional narrowing of its width. This behavior results from the fact that at the peak region the system is forced into an unfavorable interlayer stacking, whereas at the flat regions the layers assume nearly optimal stacking (see SM [26] section 1). Hence, the reduction of the peak height at larger external loads requires a higher energy penalty as compared to the flat regions.

These findings can now be used to construct a simplistic model, based on geometric arguments, that captures all physical characteristics required to reproduce the NFCs phenomenon. We recall that the frictional response is dominated by energy dissipation due to the out-of-plane motion of the graphene layer. The corresponding vertical velocity of the carbon atoms due to the ridge soliton-like motion can be estimated as $v_z^{\max} \approx \Delta z/\Delta t$, where $\Delta z$ is the moiré ridge height and $\Delta t$ is the typical passage time of the domain-wall during sliding. The latter can be estimated as $\Delta t \approx \Delta x/2v_m$, where $\Delta x/2$ and $v_m$ are half of the ridge width at half-height and the sliding velocity of the moiré pattern, respectively. To evaluate $v_m$, we note that if the graphene layer is displaced by one substrate lattice spacing, $a_{h-BN}$, the final configuration is undistinguishable from the initial one, hence, the moiré ridge must have moved by $\lambda_m$. Therefore, the superstructure propagation velocity is given by $v_m \approx \frac{\lambda_m}{a_{h-BN}} v_{x,com} \approx 50 v_{x,com}$, where $v_{x,com}$ is the velocity of the center-of-mass of the graphene layer in the sliding direction. Finally, the frictional power in equation (5) can be estimated as $m_C \eta_z (v_z^{\max})^2 = 4 m_C \left(\frac{\lambda_m v_{x,com}}{a_{h-BN}}\right)^2 \eta_z (\Delta z/\Delta x)^2$. Figure 4c presents the load dependence of this expression using the height and width variations appearing in Fig. 4b. To obtain a dimensionless parameter, we normalize the results by the characteristic power dissipated via the translational degree of freedom in the sliding direction $m_C \eta_x v_{stage}^2$. Clearly, the simplistic model reproduces well the non-monotonic load-dependence of the friction force leading to the NFCs, including an overall 20 % frictional reduction between 0 and ~5 GPa. We note that in the range of high normal loads (above the turnover point) the simplistic model exhibits somewhat larger frictional increase than direct atomistic simulations. This can be attributed to the fact that the variations of the characteristic ridge height become smaller for increasing normal load thus affecting the accuracy of the model.

The qualitative explanation given by this model is further supported by our simulations. Figure 4d shows a typical vertical velocity temporal profile of a carbon atom within the graphene layer. The average magnitude of the vertical velocity first reduces with the increase of pressure from 0 to 4 GPa. However, further increase of the external pressure to 11 GPa results in sharpening of the ridge regions, as explained above, yielding increasing vertical velocities. The corresponding time-average of the square vertical velocity $\langle(v_{z,i} - v_{z,com})^2\rangle$, which provides a measure of the energy dissipated via out-of-plane atomic motion, correlates well (see inset in Fig. 4d) with the non-monotonic behavior of the friction-force presented above.

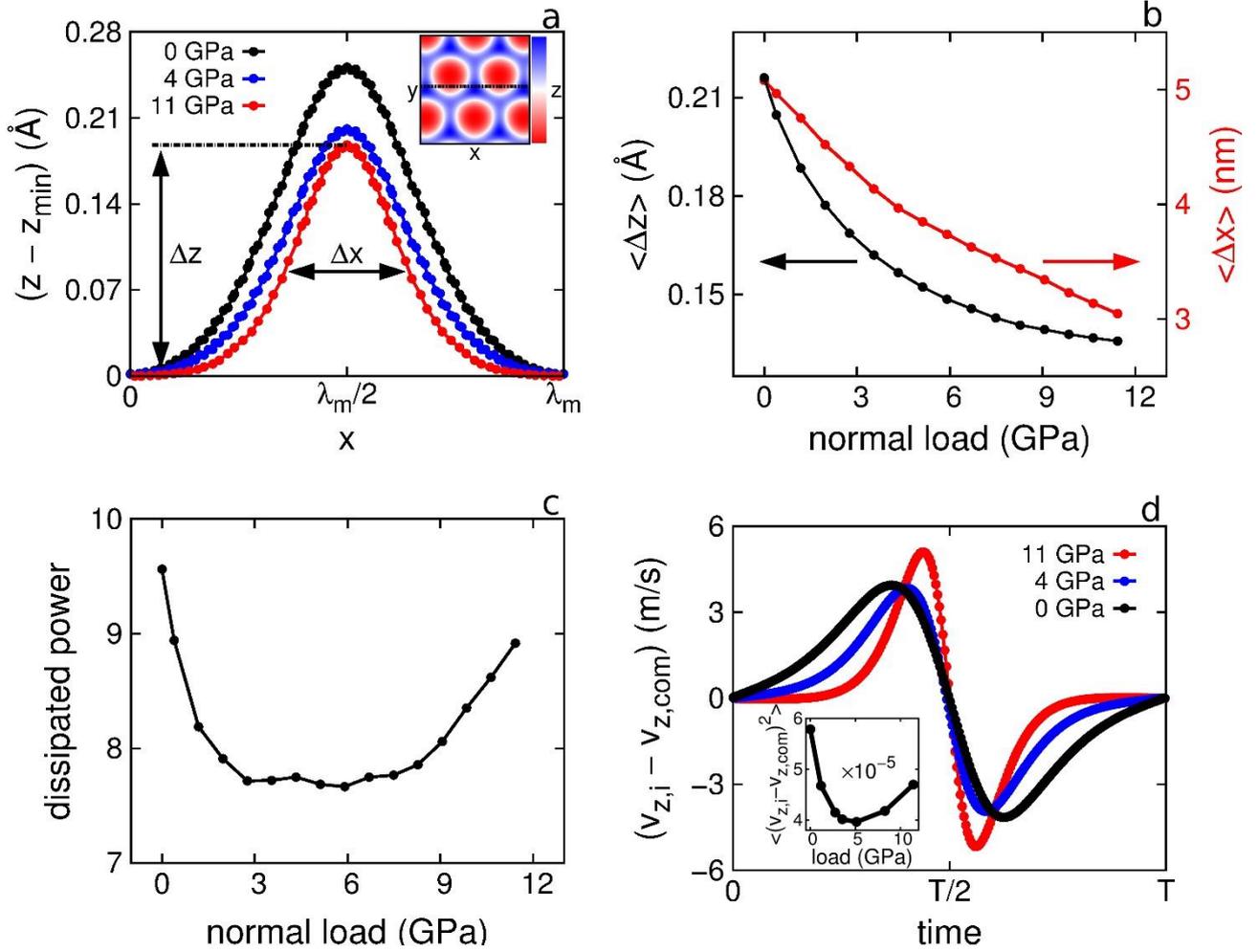

**Figure 4**: Mechanism of NFCs. (a) The moiré superstructure ridge cross-section at an applied pressure of 0 (black), 4 (blue), and 11 (red) GPa. The scan-line (see inset) corresponds to the maximum peak-dip value of the vertical deformations of graphene along the direction of sliding, $x$, expressed in units of the moiré periodicity of $\lambda_m \approx 14$ nm. The definitions of the maximum height, $\Delta z$, and the half-width-at-half-maximum, $\Delta x$, are marked by the double-headed arrows. (b) Load dependence of the characteristic peak-dip value $\langle \Delta z \rangle$ (black) and full-width-at-half-maximum $\langle \Delta x \rangle$ (red) of the moiré ridge. (c) Load dependence of the normalized power dissipated via vertical motion obtained from the simplistic model (see SM [26] section 7 for further details). (d) Vertical velocity of the carbon atom, which displays the largest velocity fluctuations during sliding under applied normal pressures of 0 (black), 4 (blue), and 11 (red) GPa, measured within the center-of-mass reference frame. Time is expressed in units of the period of the trajectory at steady-state, $T = a_{h-BN}/v_{stage} \approx 25$ ps. Inset: load dependence of the time-averaged value of $\langle (v_{z,i} - v_{z,com})^2 \rangle$, proportional to the energy dissipation, where $v_{z,i}$ is the given atom vertical velocity.

Therefore, we now have a simple and clear understanding of the physical mechanism underlying the NFCs predicted to appear in extended monocrystalline graphite/*h*-BN heterojunctions. Out-of-plane atomic motion plays a key role in this effect. Specifically, the competition between height reduction and narrowing of the moiré superstructure ridges under external pressure leads to an initial reduction of the magnitude of the slider atoms vertical velocities followed by their subsequent increase. This leads to a non-monotonic load dependence of the interfacial frictional dissipation and its unusual temperature dependence. Notably, recent theoretical predictions and experimental observations reported NFCs for atomic force microscopy tips sliding atop graphitic surfaces [37–40]. However, the origin there was identified as partial delamination of the substrate under tip retraction [37] or the reduction of surface puckering with increasing load in suspended systems [38–40]. Furthermore, a recent computational study predicted NFCs in homogeneous monocrystalline $MoS_2$ contacts due to flattening of the sliding potential energy surface under extremely high loads [41]. The unique mechanism that we predict herein is of general nature, emerges at moderate external loads, and should apply to many other layered materials heterojunctions when placed in an aligned configuration. With this respect, increasing the interfacial misfit angle results in rapid reduction of the amplitude of the out-of-plane corrugation, therefore the predicted NFCs effect is expected to become less pronounced. We can therefore conclude that in order to achieve NFCs within the suggested mechanism the interface should: (i) be *superlubric* to avoid center-of-mass stick-slip motion; (ii) be *anisotropic* with high intralayer stiffness and low bending rigidity to allow for the formation of a corrugated superstructure (see e.g. $MoS_2$/$WSe_2$ interfaces, for which large out-of-plane distortions of up to $\Delta z \approx$ 1 Å have been predicted [42,43]); (iii) be held at an aligned configuration to achieve large-scale corrugated moiré superstructure; and (iv) exhibit *soliton-like* superstructure motion under shear stress. With this respect it is interesting to note that a moiré pattern characterized by vertical distortions of the order of $\Delta z \approx 0.1$ Å is expected to appear for small misalignment angles also in homogeneous graphitic contacts [44,45], suggesting that NFCs may be observed in twisted bilayer graphene.

ACKNOWLEDGMENTS

D. M. acknowledges the fellowship from the Sackler Center for Computational Molecular and Materials Science at Tel Aviv University, and from Tel Aviv University Center for Nanoscience and Nanotechnology. W. O. acknowledges the financial support from a fellowship program for outstanding postdoctoral researchers from China and India in Israeli Universities. M. U. acknowledges the financial support of the Israel Science Foundation, Grant No.1316/13, and of the Deutsche Forschungsgemeinschaft (DFG), Grant No. BA 1008/21-1. O. H. is grateful for the generous financial support of the Israel Science Foundation under grant no. 1586/17 and the Naomi Foundation for generous financial support via the 2017 Kadar Award. This work is supported in part by COST Action MP1303.

# Supplemental Material for

# "Negative Friction Coefficient in Superlubric Graphite/Hexagonal Boron Nitride Heterojunctions"

Davide Mandelli, Wengen Ouyang, Oded Hod,[*] Michael Urbakh

Department of Physical Chemistry, School of Chemistry, The Raymond and Beverly Sackler Faculty of Exact Sciences and The Sackler Center for Computational Molecular and Materials Science, Tel Aviv University, Tel Aviv 6997801, Israel.

In this Supplemental Material we provide additional details on the following issues:

1. Description of the Moiré Superstructure
2. Methods
3. Convergence of the Simulation Results with Respect to the Super-cell's Lateral Size
4. Finite Temperature Calculations
5. Effect of the Multi-Layer Graphene Thickness on the Vertical Distortions of the Moiré Superstructure
6. Analysis of the Frictional Dissipation
7. Evaluation of the Vertical Energy Dissipation within the Simplistic Model

# 1. Description of the Moiré Superstructure

Figures S1 and S2 report colored maps showing the out-of-plane and in-plane deformations of the graphene/*h*-BN interfacial bilayer of our model heterojunction (see Fig. 1a in the main text) after relaxation at two different normal loads of 0 and 8 GPa. We observe a moiré pattern characterized by relatively flat regions (see top and middle panels of Figure S1), where graphene and *h*-BN respectively stretch and compress to reach local lattice matching (see top and middle panels of Figure S2). These regions are separated by domain-walls consisting of elevated ridges (see top and middle panels of Figure S1), where graphene (*h*-BN) is locally compressed (expanded) (see top and middle panels of Figure S2). Comparing the results obtained at 0 and 8 GPa (left and right panels in Figures S1 and S2, respectively), it is evident that increasing the load induces an overall suppression of the moiré vertical distortions, which is accompanied by a widening of the flat domains and subsequent narrowing of the domain-walls.

The bottom panels of Figure S2 report colored maps of the local registry index (LRI) [1] calculated for the interfacial graphene/*h*-BN bilayer, where bright yellow corresponds to local realizations of the optimal C stacking-mode, whereas darker tones indicate the energetically less favorable A' (red/violet), and the least favorable A (black) stacking-modes. The LRI colored maps clearly demonstrate that the flat domains correspond to nearly-commensurate regions of energetically favorable C stacking-mode, while the elevated ridges coincide with areas of energetically unfavorable A and A' stacking-modes. Consequently, the local interlayer distance attains its largest values in correspondence of the domain-walls coinciding with realizations of the least favorable stacking mode (see bottom panels of Figure S1), owing to the enhanced Pauli repulsion exerted by the nitrogen atoms on the eclipsed carbon atoms positioned directly above. To get further insights about this aspect, in Figure S3 we report in red the load dependence of the difference between the equilibrium interlayer distances of an artificially commensurate graphene/*h*-BN bilayer at the energetically least favorable A-stacking mode and at the energetically most favorable C-stacking mode, along with the load dependence of the vertical distortions of graphene extracted from our model multilayer heterojunction (black curve). Both curves display the same qualitative behavior, characterized by a strong initial decrease which becomes milder at larger loads, thus demonstrating that the evolution with load of the moiré vertical distortions is mainly dictated by the differences in the local stacking geometry. We note here that the difference of ~0.1 Å between the values in the two curves is due to the differences of the adopted models: a bilayer in the case of the artificially commensurate junction and a mobile multilayer junction in the case of our incommensurate model graphene/*h*-BN contact. In fact, when considering a simpler incommensurate heterojunction model consisting of a graphene monolayer over a single rigid *h*-BN layer, we obtain a load dependence of the graphene vertical distortions, which is very close to the difference between the A-stacking and C-stacking equilibrium interlayer distances obtained for a commensurate bilayer (see blue curve in Figure S3).

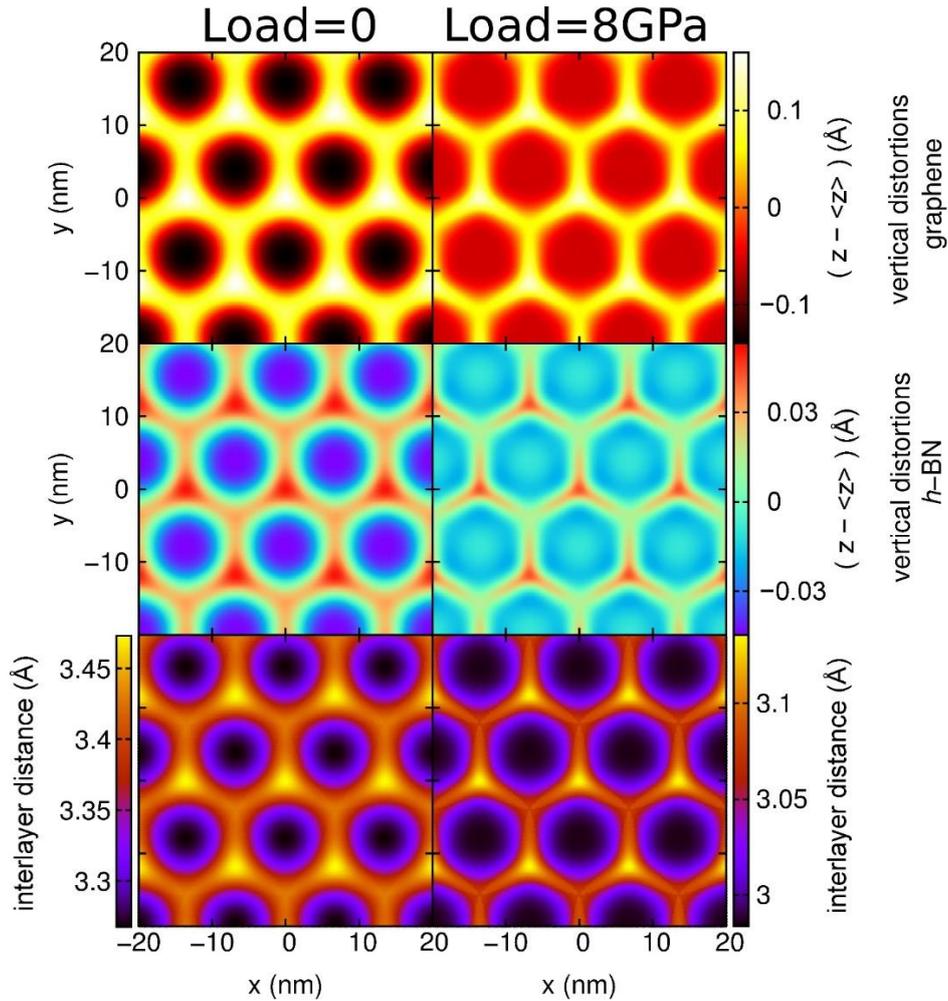

**Figure S1:** Top panels: colored maps showing the out-of-plane deformations of the graphene layer, calculated with respect to its average basal plane, when relaxed over *h*-BN at two different normal loads of 0 and 8 GPa. As the load is increased, the amplitude of the out-of-plane distortions reduces. Middle panels: colored maps of the corresponding out-of-plane deformations measured within the first *h*-BN layer in contact with graphene. The same load-induced suppression of the amplitude of the elevated ridges occurs also within the *h*-BN substrate. Bottom panels: colored maps of the local interlayer distance between graphene and *h*-BN, computed as the perpendicular distance of each carbon atom from the nearest nitrogen or boron atom within the substrate.

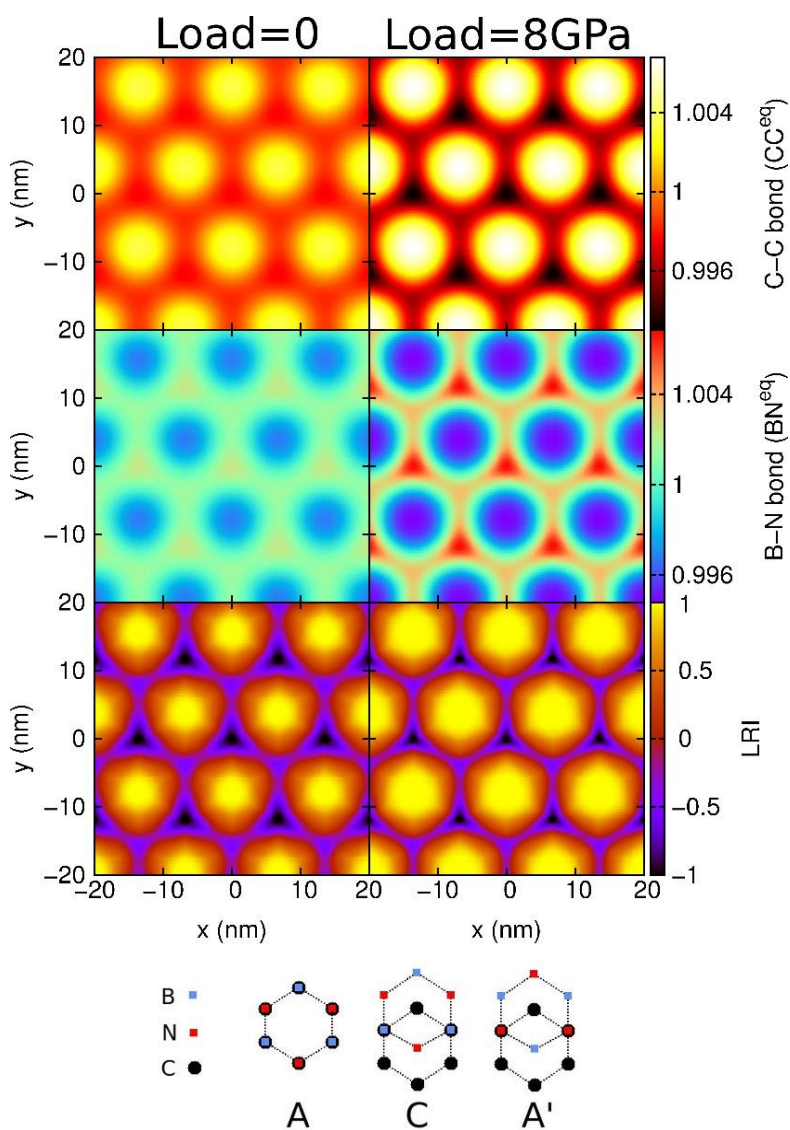

**Figure S2:** Top panels: colored maps showing the average carbon-carbon distances (in units of the REBO equilibrium value of $CC^{eq} = 1.3978$ Å. Here, four digits after the decimal point are reported for reproducibility purposes) within the graphene layer relaxed over *h*-BN at two different normal loads of 0 and 8 GPa. As the load is increased, the amplitude of the in-plane distortions increases. Middle panels: colored maps of the corresponding average boron-nitrogen distances (in units of the equilibrium value of $BN^{eq} = 1.4232$ Å. Here, four digits after the decimal point are reported for reproducibility purposes) measured within the first *h*-BN layer in contact with graphene. The same load-induced enhancement of the amplitude of the in-plane distortions occurs also within the *h*-BN substrate. Bottom panels: colored maps of the local registry index (LRI) [1] calculated for the interfacial graphene/*h*-BN bilayer, where bright yellow corresponds to local realizations of the optimal C stacking-mode, whereas darker tones indicate the energetically less favorable A' (red/violet), and the least favorable A (black) stacking-modes. As the load is increased the areas of the nearly commensurate regions of C stacking-mode increases. Several high-symmetry stacking-modes are schematically represented at the bottom of the figure.

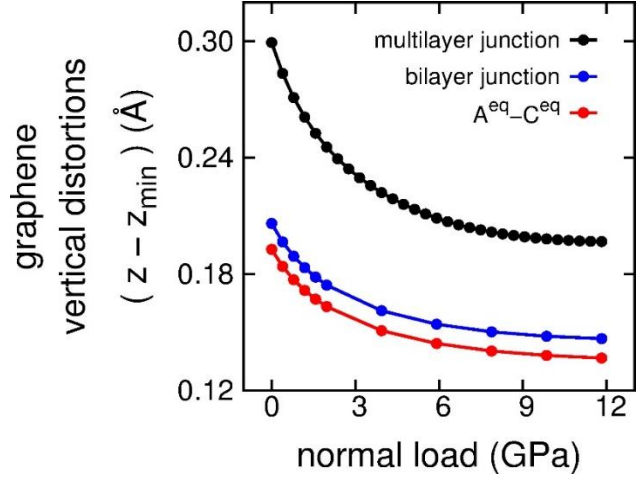

**Figure S3:** Normal load dependence of the amplitude of the vertical distortions of graphene measured in our multilayer heterojunction (black) and in a model graphene/*h*-BN bilayer with rigid *h*-BN (blue). The red curve shows the difference between the equilibrium interlayer distances of an artificially commensurate graphene/*h*-BN bilayer (at the lattice spacing of *h*-BN) at the energetically least unfavorable A-stacking mode and at the energetically most favorable C-stacking mode.

Finally, we note that the amplitude of the out-of-plane vertical distortions is found to be roughly three times smaller in *h*-BN than in graphene (see top and middle panel of Figure S1). The origin of this result lies in the different nature of the in-plane strain field characterizing the two materials in the relaxed heterojunction. Notably, both graphene and *h*-BN are practically flat within the nearly commensurate domains (see top and middle panels of Figure S1) and most of the vertical distortions are concentrated in the domain walls of the moiré superstructure. In graphene, these domain walls are characterized by intra-layer bond compression (see top panels of Figure S2), which favors out-of-plane distortions to release the excess stress. In *h*-BN these regions exhibit intra-layer bond expansion (see middle panels of Figure S2), which tends to keep the layer flat. The observed distortions of *h*-BN thus aim to optimize the interlayer interactions with the adjacent corrugated graphene layer, but their amplitude is limited by the local in-plane strain.

## 2. Methods

### 2.1. System Description

Our model interface includes a graphene monolayer sliding over a mobile *h*-BN substrate. We consider the aligned geometry, where the crystalline axes of the contacting two-dimensional (2D) crystals are parallel. Graphene and *h*-BN layers are built adopting the following 2D primitive vectors of the corresponding triangular lattices: $\mathbf{g}_1 = a_g(1,0)$, $\mathbf{g}_2 = a_g(1/2, \sqrt{3}/2)$, and $\mathbf{b}_1 = a_{h-BN}(1,0)$, $\mathbf{b}_2 = a_{h-BN}(1/2, \sqrt{3}/2)$. In order to describe the experimental ratio $\rho = a_{h-BN}/a_g \approx 1.018$ [2] between the lattice constant of *h*-BN and of graphene, we fix $a_g$, and we set $a_{h-BN} = \frac{56}{55} a_g \approx 1.0182\, a_g$. This choice allows us to build a commensurate triangular super-cell of primitive lattice vectors $\mathbf{L}_1 = 56\, \mathbf{g}_1 = 55\, \mathbf{b}_1$ and $\mathbf{L}_2 = 56\, \mathbf{g}_2 = 55\, \mathbf{b}_2$, containing $N_C = 6{,}272$ carbon atoms in the graphene layer and $N_{h-BN} = 6{,}050$ boron and nitrogen atoms in each layer of *h*-BN. All simulations discussed in the main text are performed adopting a four-layers thick *h*-BN substrate, corresponding to a total number $N_{tot} = 30{,}472$ of atoms in the super-cell. In this study, we are mainly interested in the frictional response of extended monocrystalline systems subjected to a uniform normal load. To mimic an extended contact, we implement periodic boundary conditions in the lateral directions. We further checked that the results are converged with respect to the model system super-cell dimensions, as discussed in section 3, below. Our super-cell model is therefore expected to represent well infinite and truly incommensurate interfaces.

### 2.2. Simulation Setup

In our simulation setup, aiming to mimic recent friction-force experiments [3], a graphene monolayer is dragged along a flexible *h*-BN substrate. The pulling apparatus is represented by a rigid duplicate of the dragged monolayer, positioned parallel to the underlying surface, whose center of mass is displaced along the substrate to represent the experimental moving stage [see Fig. 1(d) of the main text]. The moving stage is built as a flat graphene monolayer where the in-plane (*x*,*y*) coordinates of the carbon atoms are taken equal to those of the carbon atoms within the graphene slider *after* relaxation over *h*-BN (i.e., strained graphene). Normal loads are modeled by applying a uniform constant force to each atom of the dragged monolayer.

The in-plane interactions between the rigid duplicate and the sliding graphene monolayer are described by lateral harmonic springs connecting each graphene atom with its counterpart on the rigid duplicate. The presented results have been obtained using a spring constant of $k_\parallel = 11$ meV/Å$^2$ in the lateral $(x, y)$ directions for all carbon atoms. This value corresponds to the equilibrium curvature of the Kolmogorov-Crespi [4] potential for lateral displacements at the equilibrium interlayer distance (3.37 Å) of a fully relaxed graphene bilayer. The interlayer interactions between the graphene monolayer and the *h*-BN substrate, and between different *h*-BN layers are described via the graphene/*h*-BN heterogeneous interlayer potential (g*h*-ILP) [5] and by the *h*-BN/*h*-BN homogeneous interlayer potential (*hh*-ILP) [6], respectively. The intralayer interactions within the graphene monolayer are described using the REBO potential [7,8]. The intralayer interactions within each *h*-BN monolayer are computed via the Tersoff potential, as parameterized in Ref. [9]. We note that the interlayer potentials were designed to augment intralayer terms. To this end, each atom is assigned a layer identifier such that the interactions between atoms residing on the same layer are described by the intralayer term, whereas the interactions between atoms residing on different layers are described by the interlayer term.

The equilibrium carbon-carbon distance of the adopted REBO potential [7,8] is $CC_{REBO} = 1.3978$ Å. This value is used to construct the graphene monolayer, whereas for *h*-BN we use a boron-nitrogen equilibrium distance of $BN_{super-cell} = \frac{56}{55} CC_{REBO} \approx 1.4232$ Å. We note that the equilibrium boron-nitrogen distance of the adopted *h*-BN intralayer Tersoff potential [9], $BN_{Tersoff} \approx 1.44$ Å, differs by ~0.02 Å from that of the commensurate super-cell. In order to avoid any residual stress, we implement a rigid shift of all distances in the Tersoff potential that allows us to tune the equilibrium lattice spacing to the desired value [see Figure S4(a)]. We checked that the elastic properties of *h*-BN remain unchanged by comparing the phonon dispersion curves computed with our "shifted" potential, to those obtained adopting the original version [see Figure S4(b)].

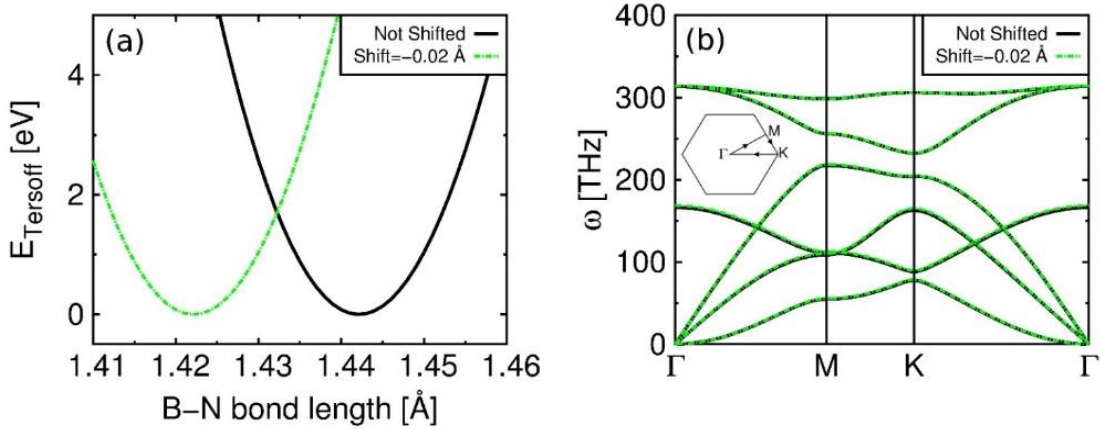

**Figure S4:** Panel (a) shows the total energy curves of single-layer *h*-BN as a function of the boron-nitrogen bond-length obtained with the original Tersoff potential of Ref. [9] (solid black curve) and after the implementation of a shift by −0.02 Å of all the distances (green dotted-dashed curve). The overall result is a rigid shift that leaves the curvature unaffected. Panel (b) is a comparison between the phonon dispersion curves of single-layer *h*-BN calculated with the two versions of the intralayer potential, along the path in the Brillouin zone schematically reported in the inset. Differences are found to be negligibly small.

### 2.3. Simulation Protocol

The starting configurations for the sliding simulations are generated as follows. For each applied normal load we first perform a full geometry relaxation of the model interface using the FIRE algorithm [10], in absence of the rigid stage. In all simulations all atoms are free to move in any (*x*,*y*,*z*) directions, apart from those residing in the bottommost *h*-BN layer, which are held fixed at their equilibrium positions and those of the graphene duplicate that are rigidly shifted along the *x* direction at a constant velocity. The starting interlayer distance between graphene and *h*-BN, and between the four *h*-BN layers is of 3.34 Å. The relaxation procedure is terminated when the forces acting on each degree of freedom reduce below $10^{-6}$ eV/Å. Starting from the fully relaxed configurations, sliding simulations are performed by moving the rigid duplicate at a constant velocity of $v_{\text{stage}} = 10$ m/s along the zigzag direction of the substrate, corresponding to the *x*-axis of our Cartesian reference frame.

All zero temperature (*T* = 0 K) simulation results presented in the main text are obtained by solving the following equations of motion:

$$m_C \ddot{r}_i = -\nabla_{r_i}\left(V_{\text{inter}}^{gh-\text{ILP}} + V_{\text{intra}}^{\text{REBO}}\right) + k_\parallel\left(r_{\parallel,i}^{\text{stage}} - r_{\parallel,i}\right) - F_N \hat{z} + F_i^{\text{damp}}, \quad (1)$$

for the carbon atoms within the graphene slider, and

$$m_{B,N} \ddot{r}_i = -\nabla_{r_i}\left(V_{\text{inter}}^{gh-\text{ILP}} + V_{\text{inter}}^{hh-\text{ILP}} + V_{\text{intra}}^{\text{Tersoff}}\right) + F_i^{\text{damp}}, \quad (2)$$

for the nitrogen and boron atoms within the substrate, respectively. Here $m_{C,B,N}$ are the atomic masses of carbon, boron, and nitrogen. The first two terms in the r.h.s. of equation (1) and the first three terms in the r.h.s. of equation (2) are the forces acting on the carbon atoms and on the substrate boron and nitrogen atoms due to the intra- and interlayer interactions. The third term in equation (1) is the elastic driving acting in the $(x, y)$ plane due to the springs of constant $k_\parallel$ connecting each carbon atom to its counterpart in the moving rigid stage, while the fourth term is a homogeneous force acting in the vertical direction, mimicking an external normal load. To account for energy dissipation, we introduce viscous forces acting directly on each atom:

$$F_i^{\text{damp}} = -m_{C,B,N} \sum_{\alpha=x,y,z} \eta_\alpha v_{\alpha,i} \hat{\alpha}. \quad (3)$$

Here $\hat{\alpha}$ is a unit vector along direction $\alpha = x, y, z$, $v_{\alpha,i}$ is the velocity component of atom $i$ along the $\alpha$ direction, and $\eta_\alpha$ is the corresponding damping coefficient, required to reach a steady-state motion [11]. In all simulations, we fix the damping anisotropy to $\eta_z/\eta_{x,y} = 45$. For the adopted interface model with four substrate layers, this ratio was shown to reproduce well the experimental frictional angular anisotropy of twisted graphite/h-BN heterojunctions [3]. In practice, we use $\eta_{x,y} = 0.1$ ps$^{-1}$ and $\eta_z = 4.5$ ps$^{-1}$, which fall within the typical range adopted in molecular dynamics simulations of nanoscale friction [11], and are close to theoretical estimations of damping coefficients for surface atomic adsorbates [12] (see section 2.4 below for further details regarding this choice of the damping coefficients). The equations of motions are solved using a velocity-Verlet integrator and a time step of 1 fs, which was previously found to be sufficiently small to achieve numerical stability in molecular dynamic simulations of similar systems using the same protocol [13].

To investigate the effects of finite temperature, we performed Langevin dynamics simulations by including in equations (1), (2) a random force satisfying the fluctuation-dissipation theorem. A detailed discussion of the results and of the protocol adopted in these simulations is reported in section 4, below.

## 2.4. Evaluation of Frictional Dissipation

In our simulations, the instantaneous shear-force is given by the sum of the lateral spring forces, $F_i(t) = k_\parallel \left( r_{\parallel,i}^{\text{stage}} - r_{\parallel,i} \right)$, acting between all atoms of the sliding graphene layer and their duplicates within the rigidly moving stage. The kinetic friction force, $F_k$, is evaluated as the time average of the total shear-force acting on the moving stage in the sliding direction, $F_k = \langle \sum_{i=1}^{N_C} F_{x,i}(t) \rangle$. The time average is taken after the initial transient dynamics decays and the system reaches steady-state motion (see Figure S5), corresponding to a periodic trajectory with periodicity equal to $T = \frac{a_{h-BN}}{v_{\text{stage}}} \approx 25$ ps.

At steady-state, the power $\langle P_{\text{in}}(t) \rangle$ generated by the external springs equals the power $\langle P_{\text{out}}(t) \rangle$ dissipated by the internal viscous forces. The first is given by

$$\langle P_{\text{in}}(t) \rangle = \sum_{i=1}^{N_C} \langle \mathbf{F}_i(t) \cdot \mathbf{v}_i(t) \rangle = F_k v_{\text{stage}}, \tag{4}$$

while the second can be written as the sum of contributions coming from each layer, namely, $\langle P_{\text{out}}(t) \rangle = \sum_{i=1}^{N_{\text{layers}}} \langle P_{\text{out}}^i(t) \rangle$, where

$$\langle P_{\text{out}}^i(t) \rangle = \sum_{k=1}^{N_i} m_k^i \sum_{\alpha=x,y,z} \left[ \eta_\alpha \langle \left( v_{\alpha,k}^i(t) \right)^2 \rangle \right] =$$

$$= \sum_{k=1}^{N_i} m_k^i \sum_{\alpha=x,y,z} \left[ \eta_\alpha \langle \left( v_{\alpha,k}^i(t) - v_{\alpha,\text{com}}^i(t) \right)^2 \rangle \right] + M_i \sum_{\alpha=x,y,z} \eta_\alpha \langle \left( v_{\alpha,\text{com}}^i(t) \right)^2 \rangle. \tag{5}$$

Here $N_i$, $\mathbf{v}_{\text{com}}^i$ and $M_i$ are the total number of atoms in the $i$-th layer, its center-of-mass velocity and its total mass, respectively. Equating (4) and (5) leads to the following expression for the kinetic frictional stress:

$$\sigma_k = \frac{F_k}{A} = \frac{\sum_{i=1}^{N_{\text{layers}}} \left[ \sum_{k=1}^{N_i} m_k \sum_{\alpha=x,y,z} \eta_\alpha \langle \left( v_{\alpha,k}^i(t) - v_{\alpha,\text{com}}^i(t) \right)^2 \rangle + M_i \sum_{\alpha=x,y,z} \eta_\alpha \langle \left( v_{\alpha,\text{com}}^i(t) \right)^2 \rangle \right]}{A \cdot v_{\text{stage}}}, \tag{6}$$

where $A$ is the contact area.

Equation (6) demonstrates that the frictional stress can be divided into center-of-mass motion contributions in the lateral and perpendicular directions and corresponding terms related to the kinetic energy dissipated via the internal degrees of freedom. Careful analysis of the various terms allows us to explain the frictional behavior of the heterogeneous junction subjected to different external normal loads in terms of the nature of the dominating dissipative channel(s).

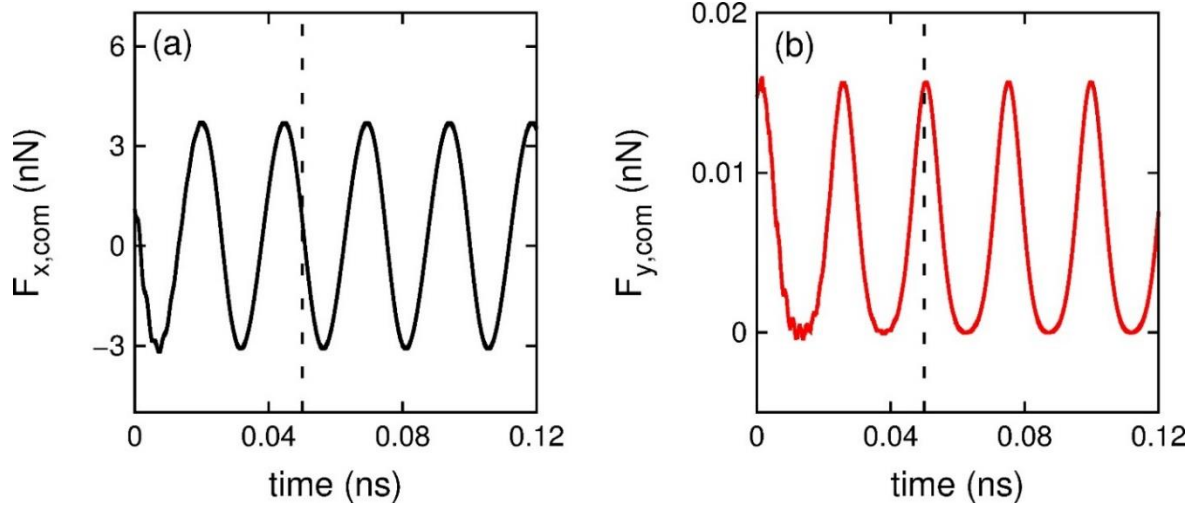

**Figure S5:** (a) The instantaneous total spring force in the *x* direction of sliding along the trajectory of the simulation at an applied normal load of 4 GPa. (b) The corresponding total spring force in the *y* direction. Averages are computed at steady-state, neglecting the initial transient as indicated by the dashed lines in panels (a) and (b).

It is worth stressing that in our simulations the absolute value of the frictional stress is controlled by the absolute value of the damping coefficients $\eta_z, \eta_{x,y}$. To check the sensitivity of our results towards the values of the damping coefficients, we performed test simulations using $\eta_z = 0.45$ ps$^{-1}$ and $\eta_{x,y} = 0.01$ ps$^{-1}$, one order of magnitude smaller with respect to the values of $\eta_z = 4.5$ ps$^{-1}$, $\eta_{x,y} = 0.1$ ps$^{-1}$ adopted for the simulations discussed in the main text. The results, presented in Figure S6, show that both the averaged square center-of-mass velocities $\left\{\left(v_{\alpha,\text{com}}^i(t)\right)^2\right\}$, and the distributions of the atomic velocities in the center-of-mass frame of reference $\{(v_{\alpha,k}^i(t) - v_{\alpha,\text{com}}^i(t))\}$ are left practically unchanged. Within this regime, it follows directly from equation (6) that any relative change of the frictional stress as a function of the applied normal load is independent of the adopted absolute values of the damping coefficients and depends only on their ratio $\eta_z/\eta_{x,y}$. For the latter, we used a value of $\frac{\eta_z}{\eta_{x,y}} = 45$, which was previously fitted to reproduce the ~4-fold kinetic friction angular anisotropy of twisted graphite/*h*-BN contacts [3].

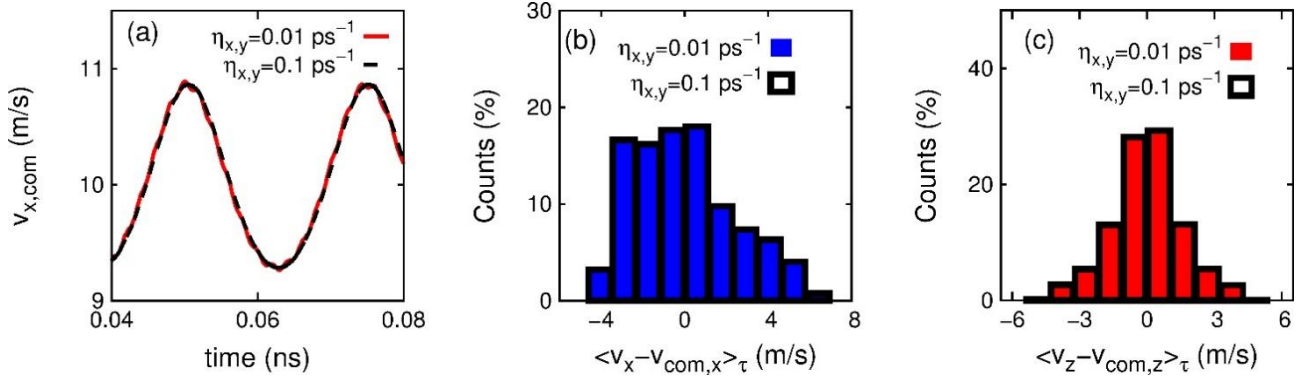

**Figure S6:** (a) Center-of-mass velocity in the direction of sliding of the graphene layer as a function of time, under an applied normal load of 4 GPa. The continuous red curve and the black dashed curve correspond to simulations performed adopting damping coefficients of $\eta_z = 0.45$ ps$^{-1}$, $\eta_{x,y} = 0.01$ ps$^{-1}$ and of $\eta_z = 4.5$ ps$^{-1}$, $\eta_{x,y} = 0.1$ ps$^{-1}$, respectively. Panels (b) and (c) report the corresponding time-averaged distributions of the velocities of the carbon atoms in the $x$ and $z$ directions, measured in the center-of-mass frame of reference, for the two different sets of damping coefficients. The results demonstrate that the simulations are performed in a regime where the averaged square value of the center-of-mass velocity and the carbon atoms velocity distributions are independent of the absolute values of the damping coefficients.

## 3. Convergence of the Simulation Results with Respect to the Super-cell's Lateral Size

The characteristic length-scale of the heterogeneous contacts is set by the periodicity $\lambda_m$ (~14 nm for the aligned case considered herein) of the moiré superstructure. The size of the simulated super-cell should therefore be sufficiently larger than $\lambda_m$ for the calculated static and dynamical properties of the interface to be converged. To check for super-cell size effects, we performed tests adopting two different super-cells of increasing size $L \approx 14$ nm $\sim \lambda_m$ and $\approx 28$ nm $\sim 2\lambda_m$, accommodating one and four moiré primitive cells, respectively. Both super-cells consist of one rigid $h$-BN substrate layer and one mobile graphene layer. The total number of carbon atoms in the cells are 6,272 and 24,642, respectively. We compare the surface corrugation at zero applied normal load (defined as the maximal amplitude of vertical carbon atoms displacements in the relaxed structure) and the corresponding kinetic frictional stress under sliding at constant velocity of 10 m/s, at zero temperature ($T = 0$ K). For both properties, we observed only minor deviations (~0.2 %) between the results obtained adopting the smaller and larger super-cell, indicating that finite-size effects are sufficiently small in the adopted simulation super-cell of dimension $L = \lambda_m$.

## 4. Finite Temperature Calculations

In the main text, we reported results obtained from simulations conducted at zero ($T = 0$ K), and room temperature ($T = 300$ K). For the latter, we adopted a standard Langevin approach for the same model system including a graphene layer sliding over a four-layers thick $h$-BN substrate, the lowest of which serves as a rigid support.

The equations of motions of the carbon atoms within the slider were modified as follows:

$$m_C \ddot{\mathbf{r}}_i = -\nabla_{\mathbf{r}_i}\left(V_{\text{inter}}^{gh-\text{ILP}} + V_{\text{intra}}^{\text{REBO}}\right) + k_\parallel\left(\mathbf{r}_{\parallel,i}^{\text{stage}} - \mathbf{r}_{\parallel,i}\right) - F_N \hat{z} + \sum_{\alpha=x,y,z}\left(\zeta_{\alpha,C} R(t) - m_C \eta_\alpha v_{\alpha,i}\right)\hat{\boldsymbol{\alpha}}, \quad (7)$$

where $R(t)$ is a delta-correlated stationary Gaussian process, namely $\langle R(t) \rangle = 0$, and $\langle R(t)R(t') \rangle = \delta(t - t')$, and the coefficients $\zeta_{\alpha,C} = \sqrt{2 m_C \eta_\alpha k_B T}$ satisfy the fluctuation-dissipation theorem. Similarly, the equations of motion of the boron and nitrogen atoms in the topmost three layers of $h$-BN were given by:

$$m_{B,N} \ddot{\mathbf{r}}_i = -\nabla_{\mathbf{r}_i}\left(V_{\text{inter}}^{gh-\text{ILP}} + V_{\text{inter}}^{hh-\text{ILP}} + V_{\text{intra}}^{\text{Tersoff}}\right) + \sum_{\alpha=x,y,z}\left(\zeta_{\alpha,B,N} R(t) - m_{B,N} \eta_\alpha v_{\alpha,i}\right)\hat{\boldsymbol{\alpha}}. \quad (8)$$

with $\zeta_{\alpha,B,N} = \sqrt{2 m_{B,N} \eta_\alpha k_B T}$. As mentioned above, the bottommost $h$-BN layer is held fixed in its equilibrium crystalline configuration.

The frictional stress at finite temperature is computed adopting the following protocol. First, we equilibrated the interface in presence of the rigid support, at rest. The latter is then set into motion at constant velocity for a total simulation time of ~8 ns. The average frictional stress is computed at steady state, neglecting the initial ~3 ns of the trajectory (see section 2.4 above for the definition of the frictional stress). All simulation parameters were the same as those used for performing the simulations at zero temperature. The equations of motion were solved using a velocity-Verlet integrator and a time step of 1 fs.

Figure S7 shows the instantaneous temperature of the interface, calculated from the kinetic energy per atom ($\varepsilon_{\text{ke}}$) as $T = 2\varepsilon_{\text{ke}}/3k_B$, along the sliding trajectory of the interface at a thermostat temperature of 300 K, and two different normal loads of 0 and 8 GPa. The instantaneous temperatures (full black and red lines) fluctuate around the thermostat temperature (dashed green line) indicating the validity of the used procedure. In Figure S8(a), (b) we report, for comparison, the steady-state friction traces obtained in simulations performed at zero temperature and ambient temperature, respectively, for three different normal loads of 0 (black curves), 4 (blue curves), and 8 GPa (red curves). At zero temperature, the instantaneous shear-stress exhibits a periodic oscillatory behavior. The amplitude of these oscillations increases with the applied normal load. At room temperature, thermal fluctuations somewhat mask the periodicity, and the instantaneous shear-stress exhibits larger fluctuations.

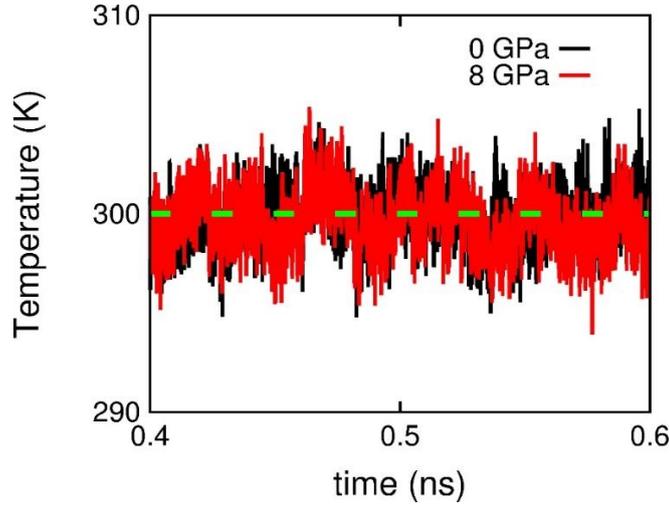

**Figure S7:** Instantaneous temperature in the sliding simulations at room temperature for two different normal loads of 0 (black curve) and 8 GPa (red curve).

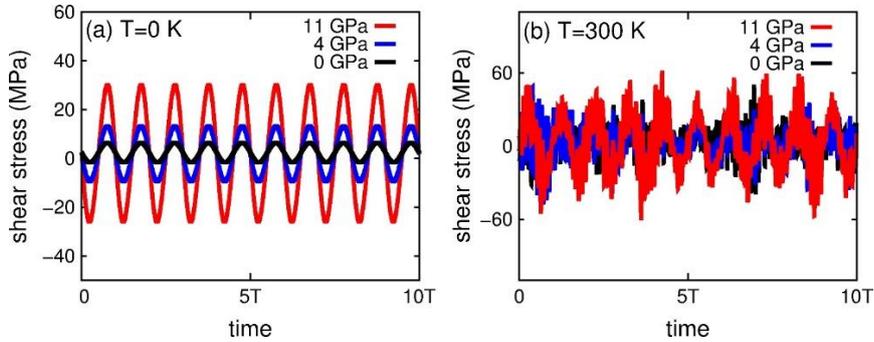

**Figure S8:** (a) The friction trace obtained in sliding simulations at zero temperature for three different normal loads of 0 (black curve), 4 (blue curve), and 8 GPa (red curve), showing a periodic behavior. Time is reported in units of the period $T = \frac{a_{h-\text{BN}}}{v_{\text{stage}}} \approx 25$ ps, corresponding to a displacement of the stage that equals the lattice spacing of the $h$-BN substrate. (b) The friction trace obtained in sliding simulations at room temperature for the same set of loads. The instantaneous shear stress exhibits larger fluctuation, nevertheless, under high loads the periodicity remains noticeable.

## 5. Effect of the Multi-Layer Graphene Thickness on the Vertical Distortions of the Moiré Superstructure

In the main text we reported results obtained considering a single graphene layer sliding over a thick *h*-BN substrate. There, we showed that the load dependence of the dissipative dynamics of the interface is mainly controlled by the amplitude of the vertical distortions of the interfacial moiré superstructure. In this section, we report results of geometry optimizations and demonstrate the suppression with load of the vertical moiré deformations for the case of a multilayer-graphene slider. The obtained results indicate the validity of the single graphene layer model adopted in the main text.

To study the dependence of the vertical distortions of the interfacial moiré superstructure on the number of layers in the graphite slider, we considered two model heterojunctions. The first consists of a graphene monolayer over a six-layers thick *h*-BN substrate. We will refer to this as the 1L/6L model, which contains $N_{tot} = 42,572$ atoms [see Figure S9(a)]. The second consists of a six-layers thick graphite slab over the same six-layers thick *h*-BN substrate. We will refer to this as the 6L/6L model, which contains $N_{tot} = 73,932$ atoms [see Figure S9(b)]. The super-cells were built following the procedure outlined in sections 2.1 and 2.2 above. Geometry optimizations were performed following the protocol reported in section 2.3 above. We further checked that simulations performed adopting a recently refined version of our interlayer potentials [13] yield qualitatively similar results. In Figure S9(c), (d) we show the super-cells of the 1L/6L and of the 6L/6L models, respectively, after geometry optimization at zero normal load, where atoms are colored according to their vertical position relative to the average basal plane of the corresponding layer.

In Figure S9(e), (f) we plot the peak-dip value of the vertical distortions of each layer of the 1L/6L and of the 6L/6L models, respectively, for two different normal loads of 0 and 6 GPa. At zero applied normal load, the amplitude of the vertical distortions within the multilayer graphene slab of the 6L/6L model is fairly constant and equal to an average value of 0.385 Å, very close to the value of 0.378 Å measured in the graphene monolayer of the 1L/6L model. Upon increasing the normal load to 6 GPa, the amplitude of the vertical distortions of graphene reduces to 0.274 Å in the 1L/6L model and to a somewhat larger average value of 0.316 Å in the 6L/6L model, corresponding to a relative reduction of 28% and 18%, respectively. This result suggests that, under sliding, for the case of a multilayer graphene slider one would obtain the same relative frictional reduction at a slightly higher normal load compared to the single graphene layer model.

Finally, we note that the peak-dip value of ~0.4 Å of the vertical distortions of the graphene layers of both the 1L/6L and 6L/6L models measured at 0 GPa, is larger than the corresponding value of ~0.3 Å obtained for the model with four layers of *h*-BN considered in the main text (see black curve in Figure S3). This further suggests that, in the case of much thicker interfaces, where the amplitude of the vertical distortions is at convergence, as expected in typical experimental setups [2], more pronounced frictional reduction could be achieved than the one predicted in the simulations reported herein.

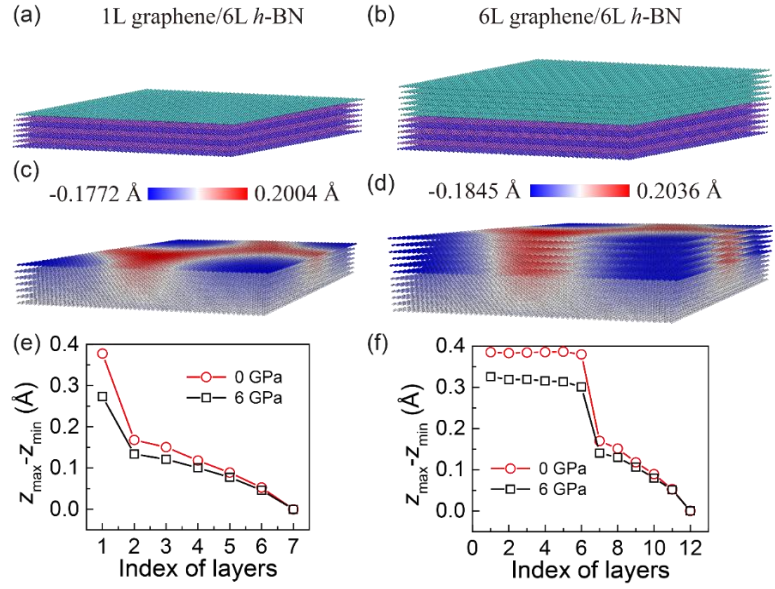

**Figure S9:** (a) The super-cell of the model heterojunction formed by a single graphene layer on top of a six-layers thick *h*-BN substrate (1L/6L model). (b) The super-cell of the model heterojunction formed by a six-layers thick graphite slab on top of a six-layers thick *h*-BN substrate (6L/6L model). Carbon atoms are colored in cyan, while boron and nitrogen atoms are colored in blue and pink, respectively. Panels (c) and (d) report the 1L/6L and 6L/6L super-cell models after geometry optimization at zero applied normal load, respectively. Atoms are colored according to their vertical position measured with respect to the average basal plane of their layer. Note that the range of the scale bar is practically the same in both models. Panels (e) and (f) show the peak-dip value of the vertical distortions for each layer of the 1L/6L and 6L/6L models, respectively, for two different normal loads of 0 (red) and ~6 GPa (black). In panel (e), layer index 1 corresponds to graphene, while layer indices from 2 to 7 correspond to *h*-BN, going from the topmost layer to the bottommost one, which was kept rigidly flat during optimization. In panel (f), layer indices from 1 to 6 correspond to graphite, going from the topmost layer to the bottommost one, while layer indices from 7 to 12 correspond to *h*-BN, going from the topmost *h*-BN layer to the bottommost one, which was kept rigidly flat during optimization.

## 6. Analysis of the Frictional Dissipation

In the main text we discussed the various contributions to the overall frictional power dissipated within the interface and interpreted the results based only on the dissipative sliding of the dragged graphene monolayer. To justify this analysis, we report in Figure S10 the layer-resolved center-of-mass contribution [panel (a) in Figure S10], internal contribution [panel (b) in Figure S10], and overall frictional power [panel (c) in Figure S10] for two different normal loads of 0 and 4 GPa (see section 2.4 above for the definition of the various contributions). Clearly, the dissipation is peaked within the graphene slider, whose contribution amounts to ~90 % of the total dissipated power. This demonstrates that the frictional response of the contact is indeed dominated by the dissipation occurring within the graphene slider, with only minor contributions coming from the $h$-BN substrate layers.

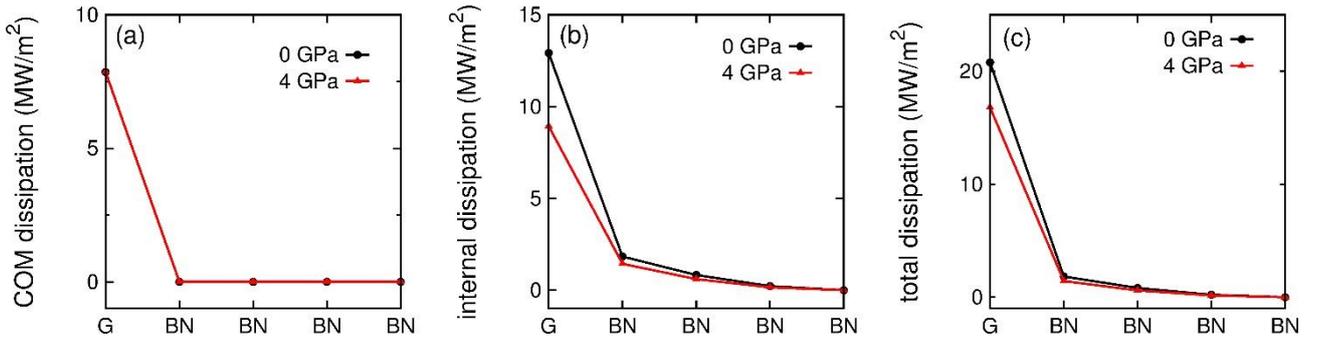

**Figure S10:** Panels (a), (b) and (c) show the center-of-mass, internal, and total frictional power dissipated in each layer of the model interface. The graphene layer and the $h$-BN layers are labeled by G and BN along the $x$-axis. Black and red curves are results from simulations at two different normal loads of 0 and 4 GPa, respectively. Dissipation is sharply peaked in the graphene layer and rapidly decreases within the substrate. We remind here that the bottommost $h$-BN layer (corresponding to the rightmost point in each panels) is rigid and does not contribute to the frictional power.

## 7. Evaluation of the Vertical Energy Dissipation within the Simplistic Model

In the main text a dimensionless parameter, $\eta_z (v_z^{\max})^2 / \eta_x v_{\text{stage}}^2$, was introduced that roughly evaluates the dissipation via the out-of-plane motion of the carbon atoms of the graphene slider. The key quantity is the maximal amplitude, $v_z^{\max}$, of the velocity fluctuations along the vertical direction, which, in turn, is estimated as:

$$v_z^{\max} = 2\lambda_{\text{m}} \langle \Delta z \rangle \frac{v_{x,\text{com}}}{a_{h-\text{BN}} \langle \Delta x \rangle}. \tag{9}$$

Here, $\lambda_m \approx 14$ nm is the periodicity of the moiré superstructure, $a_{h-BN} \approx 2.47$ Å is the lattice spacing of h-BN, $\langle \Delta z \rangle$ and $\langle \Delta x \rangle$ are the characteristic height and full-width-at-half-maximum (FWHM) of the out-of-plane distortions displayed by the graphene layer, and $v_{x,com}$ is the characteristic velocity of the center-of-mass of the graphene layer. $\langle \Delta z \rangle$ and $\langle \Delta x \rangle$ were evaluated as follows. For a given normal load, we considered the fully relaxed interface at rest and extracted the vertical moiré ridge profile of the graphene layer along several scan lines running parallel to the sliding direction, as depicted in Figure S11(a). The scanned region has been chosen to include the carbon atoms that undergo the largest out-of-plane fluctuations under sliding and are therefore the ones that contribute the most to the dissipated frictional power. We considered typically 40 scan-lines along the $y$ axis, at intervals of $\Delta y = 2$ Å. An example of the results of this analysis is presented in Figure S11(b). The average profile is then used to extract the characteristic height, $\langle \Delta z \rangle$, and the characteristic FWHM, $\langle \Delta x \rangle$, as shown in Figure S11(c).

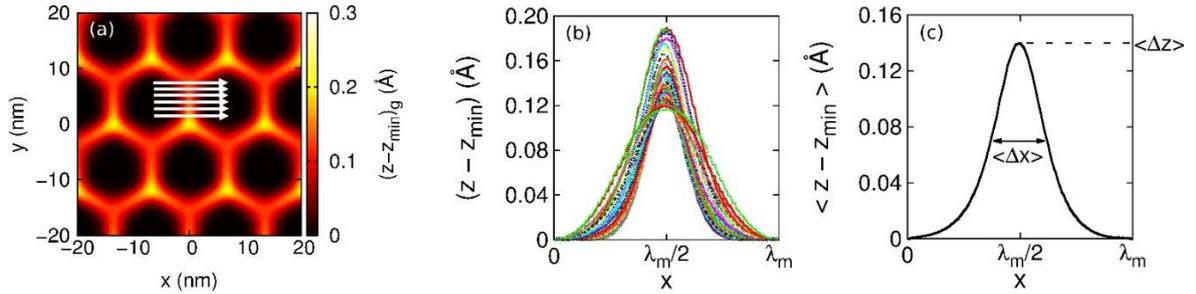

**Figure S11:** (a) Scan lines adopted to extract the height profile of the graphene out-of-plane distortions. (b) Example of the height profiles extracted along several scan lines. The position along $x$ is given in units of the moiré periodicity $\lambda_m$. (c) The average height profile calculated from the curves reported in panel (b). The definition of the characteristic height $\langle \Delta z \rangle$ and of the characteristic full-width-at-half-maximum $\langle \Delta x \rangle$ are shown. Data reported here correspond to an applied normal load of ~8 GPa.

In Figure S12 we show that the largest fluctuations of the $z$ component of the velocity of the carbon atoms coincide with the minimum of the center-of-mass velocity of the graphene layer in the $x$ direction of sliding during one period of the steady-state trajectory. Using this minimum value of $v_{x,com}$ we obtain the curve presented in Fig. 4c in the main text.

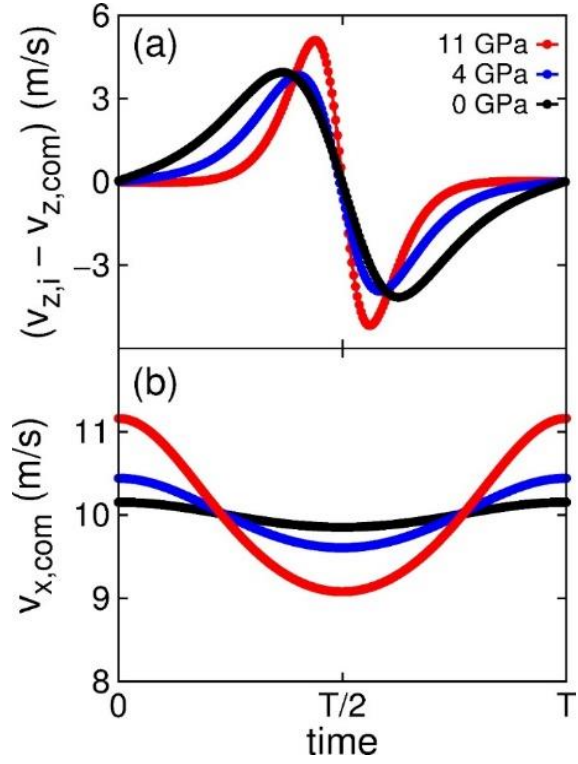

**Figure S12:** (a) Single carbon atom vertical velocity during sliding under applied normal pressures of 0 (black), 4 (blue), and 11 (red) GPa measured within the center-of-mass frame of reference (same as Fig. 4d of the main text). The chosen particle is the one displaying the largest velocity fluctuations. (b) Corresponding velocity of the center-of-mass of the slider in the $x$ direction of motion. Time is expressed in units of the periodicity of the trajectory at steady-state, $T = a_{h-BN}/v_{stage} \approx 25$ ps.